\documentclass{article}
\usepackage{times,epsfig} 

\title{Spike timing and the coding of naturalistic sounds in a central
auditory area of songbirds}

\author{Brian D. Wright,$^{1-3}$ Kamal Sen,$^{1-3}$ William
Bialek$^{1,4,5}$ and Allison J. Doupe$^{1-3}$  
 \\
$^1$Sloan--Swartz Center for Theoretical Neurobiology\\ 
$^2$Departments of Physiology and $^3$Psychiatry\\
University of California at San Francisco,
San Francisco, California 94143--0444\\
$^4$NEC Research Institute,
4 Independence Way,
Princeton, New Jersey 08540\\
$^5$Department of Physics, Princeton University, Princeton, New Jersey 08544\\
{\it
\{bdwright/kamal/ajd\}@phy.ucsf.edu, wbialek@princeton.edu}
\\ }

\begin{document}

\maketitle

\begin{abstract}

In nature, animals encounter high dimensional sensory stimuli that have complex
statistical and dynamical structure. Attempts to study the neural coding
of these natural signals face challenges both in the selection of the
signal ensemble and in the analysis of the resulting neural responses. For
zebra finches, naturalistic stimuli can be defined as sounds that they
encounter in a colony of conspecific birds. We assembled an ensemble of
these sounds by recording groups of 10-40 zebra finches, and then analyzed
the response of single neurons in the songbird central auditory area
(field L) to continuous playback of long segments from this ensemble.
   Following methods developed in the fly visual system, we measured the
information that spike trains provide about the acoustic stimulus without
any assumptions about which features of the stimulus are relevant.
Preliminary results indicate that large amounts of information are carried
by spike timing, with roughly half of the information accessible only at time
resolutions better than 10 ms; additional information is still being
revealed as time resolution is improved to 2 ms. 
Information can be
decomposed into that carried by the locking of individual spikes to the
stimulus (or modulations of spike rate) vs. that carried by timing in
spike patterns. Initial results show that in field L, temporal patterns
give at least $\sim 20$\% extra information. Thus, single central auditory 
neurons can provide an informative 
representation of naturalistic
sounds, in which spike timing may play a significant role.

\end{abstract}

\section{Introduction} 

Nearly fifty years ago, Barlow~\cite{barlow:1961} and Attneave~\cite{attneave:1954} suggested that the brain may
construct a neural code that provides an efficient representation for
the sensory stimuli that occur in the natural world. Slightly earlier,
MacKay and McCulloch~\cite{mackay_mcculloch:1952} emphasized that neurons 
that could make use of spike
timing---rather than a coarser ``rate code''---would have available a
vastly larger capacity to convey information, although they left open the
question of whether this capacity is used efficiently.  Theories for
timing codes and efficient representation have been discussed
extensively, but the evidence for these attractive ideas remains
tenuous.  A real attack on these issues requires (at least) that we
actually measure the information content and efficiency of the neural
code under stimulus conditions that approximate the natural ones.  In
practice, constructing an ensemble of ``natural'' stimuli inevitably
involves compromises, and the responses to such complex dynamic signals
can be very difficult to analyze.

At present the clearest evidence on efficiency and timing in the coding
of naturalistic stimuli comes from central invertebrate 
neurons~\cite{strong98,lewen_etal:2001}
and from the sensory periphery~\cite{rieke95,berry_etal:1997} and 
thalamus~\cite{reinagel_reid:2000,liu_etal:2001} of vertebrates.  
The situation for central
vertebrate brain areas is much less clear. 
Here we use the songbird
auditory system as an accessible test case for these ideas.  
The set of songbird telencephalic auditory areas known as the field L
complex is analogous to mammalian
auditory cortex and contains neurons that are strongly driven by natural
sounds, including the songs of birds of
the same species (conspecifics)~\cite{scheich79,lewicki96,janata99,theunissen2000}. 
We record from the zebra finch field
L, using naturalistic stimuli that consist of recordings from groups of
10-40 conspecific birds. We find that single
neurons in field L show robust and well modulated responses to playback
of long segments from this song
ensemble, and that we are able to maintain recordings of sufficient
stability to collect the large data sets that are
required for a model independent information theoretic analysis. Here we
give a preliminary account of our
experiments.

\section{A naturalistic ensemble}

Auditory processing of complex sounds is critical for perception and
communication in many species, including
humans, but surprisingly little is known about how high level brain
areas accomplish this task. Songbirds provide
a useful model for tackling this issue, because each bird within a
species produces a complex individualized
acoustic signal known as a song, which reflects some innate information
about the species' song as well as
information learned from a ``tutor" in early life. In addition to
learning their own song, birds use the acoustic
information in songs of others to identify mates and group members, to
discriminate neighbors from intruders,
and to control their living space~\cite{searcy99}. Consistent
with how ethologically critical these
functions are, songbirds have a large number of forebrain auditory areas
with strong and increasingly specialized
responses to songs~\cite{lewicki96,margoliash83,sen2001}. 
The combination of
a rich set of behaviorally relevant
stimuli and a series of high-level auditory areas responsive to those
sounds provides an opportunity to reveal
general principles of central neural encoding of complex sensory
stimuli.
Many prior studies have chosen to study neural responses to individual
songs or altered versions thereof. In order
to make the sounds studied increasingly complex and natural,
we have made recordings of the sounds
encountered by birds in our colony of zebra finches.
To generate the sound ensemble that was used in this study we first
created long records of the vocalizations of groups of 10-40 zebra
finches in a soundproof acoustic chamber with a directional microphone
above the bird cages. The group of birds generated a wide variety
of vocalizations including songs and a variety of different types of
calls. Segments of these sounds were then joined to create the sounds
presented in the experiment. One of the segments that was presented 
($\sim 30$ sec) was repeated in alternation with different segments. 

We recorded the neural responses in field L of one of the birds from the
group to the ensemble of natural sounds played back through a speaker, at
an intensity approximately equal to that in the colony
recording. This bird was lightly anesthetized with urethane. We used a 
single electrode to record the neural response
waveforms and sorted single units offline. 
Further details concerning experimental techniques can
be found in Ref.~\cite{theunissen2000}.

\section{Information in spike sequences}

\begin{figure}[t]
\centering
\includegraphics[height=4in]{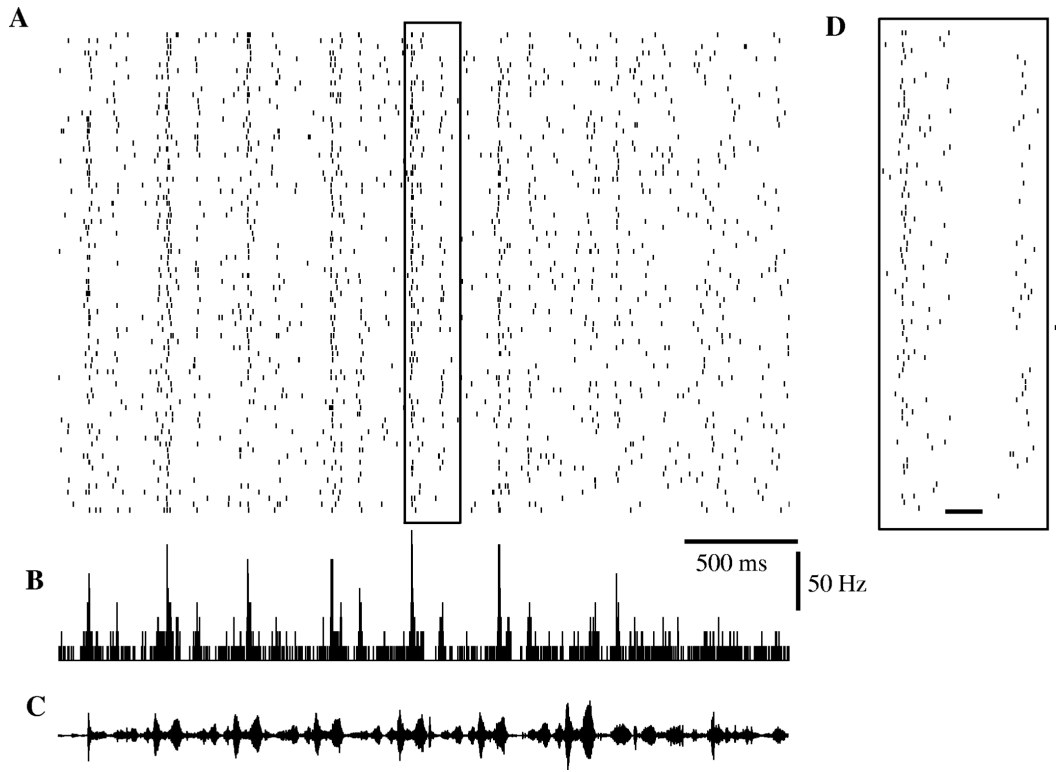}
\caption{A. Spike raster of 4 seconds of the responses of a single neuron 
in field L to a 30 second segment of a natural sound ensemble of zebra 
finch sounds. The
stimulus was repeated 80 times. B. Peri-stimulus time histogram (PSTH)
with 1 ms bins. C. Sound pressure waveform for the natural sound
ensemble. D. Blowup of segment shown in the box in A. The scale bar is
50 ms.}
\label{spikerasters}
\end{figure}

The auditory telencephalon of birds consists of a set of areas known as
the field L complex, which receive input from the auditory thalamus and
project to increasingly selective auditory areas such as NCM, cHV 
and NIf~\cite{janata99,stripling2001} and
ultimately to the brain areas specialized for the bird's own song.  Field
L neurons respond to simple stimuli such as tone bursts, and are organized
in a roughly tonotopic fashion~\cite{zaretskykonishi76}, but also respond robustly to many complex
sounds, including songs.
Figure~\ref{spikerasters} shows 4 seconds of the responses of
a cell in field L to repeated presentations of a 30 sec segment from the
natural ensemble described above.  Averaging over presentations, we see 
that spike rates are well modulated.  Looking at the responses on a finer 
time resolution we see that aspects of the spike train are reproducible 
on at least a $\sim 10$ ms
time scale. This encourages us to measure the information content of
these responses over a range of time scales, down to millisecond resolution.

Our approach to estimating the information content of spike trains
follows Ref.~\cite{strong98}.  At some time $t$ (defined relative to the
repeating stimulus) we open a window of size $T$ to look at the response. 
Within this window we discretize the spike arrival times with resolution
$\Delta\tau$ so that the response becomes a ``word'' with $T/\Delta\tau$
letters.  If the time resolution $\Delta\tau$ is very small, the allowed
letters are only 1 and 0, but as $\Delta\tau$ becomes larger one must keep
track of multiple spikes within each bin.  Examining the whole experiment,
we sample the probability distribution of words, $P_T(W)$, and the entropy
of this distribution sets the capacity of the code to convey information
about the stimulus:
\begin{equation}
S_{\rm total}(T;\Delta\tau) = -\sum_W P_T(W)\log_2 P_T(W) \,{\rm bits,}
\end{equation}
where the notation reminds us that the entropy depends both on the size of
the words that we consider and on the time resolution with which we
classify the responses. We can think of this entropy as measuring the size
of the neuron's vocabulary.

\begin{figure}[t]
\centering
\includegraphics[height=4in]{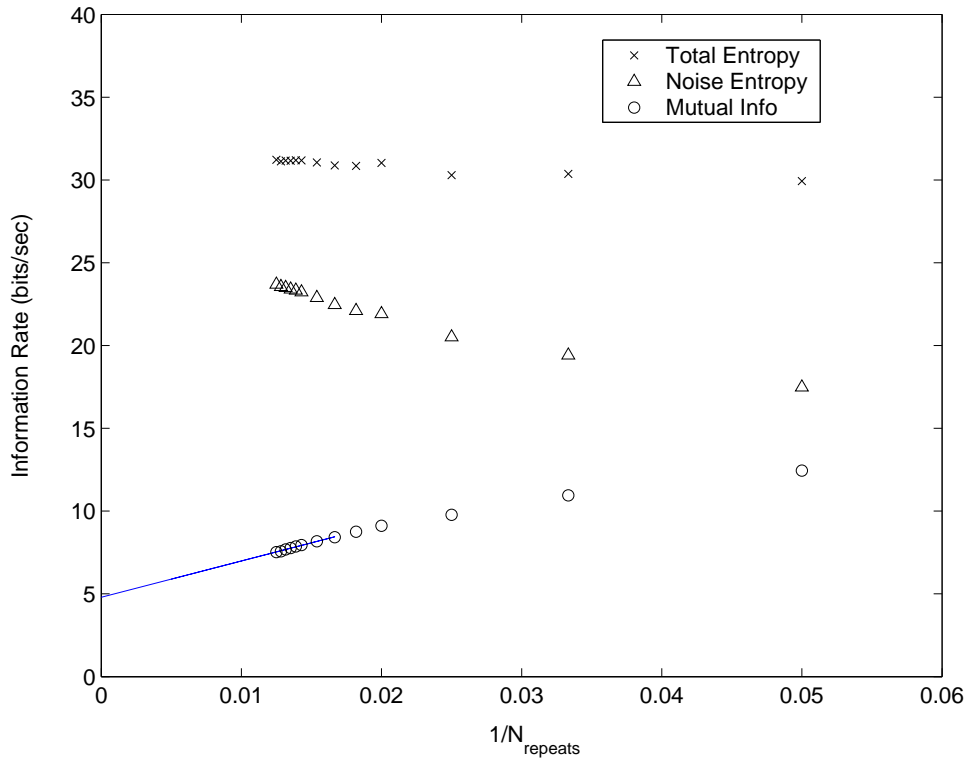}
\caption{Mutual information rate for the spike train is shown as a 
function of data size for $\Delta\tau = 2$ ms and $T = 32$ ms.}
\label{inforate_trials}
\end{figure}

Because the whole experiment contributes to defining the vocabulary size,
estimating the distribution $P_T(W)$ and hence the total entropy is not
significantly limited by the problems of finite sample size. This can be seen 
in Fig.~\ref{inforate_trials} in the stability of the total entropy with 
changing the number of repeats used in the analysis. Here we show the 
total entropy as a rate in bits per second by dividing the entropy by the 
time window $T$. 

While the capacity of the code is limited by the total entropy, to convey
information particular words in the vocabulary must be associated, more or
less reliably, with particular stimulus features.  If we look at one time
$t$ relative to the (long) stimulus, and examine the words generated on
repeated presentations, we sample the conditional distribution
$P_T(W|t)$. 
This distribution has an entropy that quantifies the noise in
the response at time $t$, and averaging over all times we obtain the
average noise entropy,
\begin{equation}
S_{\rm noise} (T;\Delta\tau) = \bigg\langle
-\sum_W P_T(W|t) \log_2 P_T(W|t) \bigg\rangle_t \,{\rm bits,}
\end{equation}
where $\langle \cdots\rangle_t$ indicates a time average (in general,
$\langle \cdots\rangle_x$ denotes an average over the variable $x$).
Technically, the above average should be an average over stimuli $s$, 
however, for a sufficiently long and rich stimulus, the ensemble average 
over $s$ can be replaced by a time average.
For the noise entropy, the problem of sampling is much more severe, since each distribution
$P_T(W|t)$ is estimated from a number of examples given by the number of
repeats.  Still, as shown in Fig.~\ref{inforate_trials}, we find that 
the dependence of our estimate on sample size is simple and regular; 
specifically, we find
\begin{equation}
S(T;\Delta\tau;N_{\rm repeats})
= S(T;\Delta\tau;\infty) + {A\over{N_{\rm repeats}}} +\cdots .
\label{extrap}
\end{equation}
This is what we expect for any entropy estimate if the distribution is 
well sampled, and if we make stronger assumptions about the sampling process 
(independence of trials etc.) we
can even estimate the correction coefficient $A$~\cite{panzeritreves95}.
In systems where much larger data sets are available this extrapolation
procedure has been checked, and the observation of a good fit to Eq.~(\ref{extrap}) is a strong indication that larger sample sizes will be
consistent with $S(T;\Delta\tau)=S(T;\Delta\tau;\infty)$; further, 
this extrapolation can be tested against
bounds on the entropy that are derived from more robust 
quantities~\cite{strong98}.  Most importantly, failure to observe 
Eq.~(\ref{extrap})
means that we are in a regime where sampling is not sufficient to draw
reliable conclusions without more sophisticated arguments, and we exclude
these regions of $T$ and $\Delta\tau$ from our discussion.

Ideally, to measure the spike train total and noise entropy rates,
we want to go to the limit of infinite word duration. 
A true entropy is extensive, which here means that it grows linearly with
spike train word duration $T$, so that the entropy rate $\mathcal{S} = S/T$ is 
constant. For finite word duration however, words sampled 
at neighboring times will have correlations between them due, in part, 
to correlations in the stimulus (for birdsong these stimulus autocorrelation 
time scales can extend up to $\sim 100$ ms). 
Since the word samples are not completely independent, the raw entropy rate 
is an overestimate of the true entropy rate. The effect is larger for smaller 
word duration and the leading dependence of the raw estimate is
\begin{equation}
\mathcal{S}(T;\Delta\tau;\infty)
= \mathcal{S}(\infty;\Delta\tau;\infty) + {B\over{T}} +\cdots ,
\label{extensextrap}
\end{equation}
where $B > 0$ and we have already taken the infinite data size limit.
We cannot directly take the large $T$ limit, since for large word 
lengths we eventually reach a data sampling limit
beyond which we are unable to reliably compute the word distributions.
On the other hand, if there is a range of $T$ for which the 
distributions are sufficiently well sampled, 
the behavior in Eq.~(\ref{extensextrap})
should be observed and can be used to extrapolate to infinite 
word size~\cite{strong98}.
We have checked that our data shows this behavior and 
that it sets in for word sizes below the limit where the data sampling 
problem occurs. For example, in the case of the noise entropy, 
for $\Delta\tau = 2$ ms, it applies for $T$ below the limit of $50$ ms 
(above this we run into sampling problems). 
The total entropy estimate is nearly perfectly extensive.

Finally, we combine estimates of total and noise entropies to obtain the
information that words carry about the sensory stimulus,
\begin{equation}
I(T;\Delta\tau) = S_{\rm total}(T;\Delta\tau) - S_{\rm noise}
(T;\Delta\tau)\,{\rm bits.}
\end{equation}
Figure~\ref{inforate_trials} shows the total and noise entropy rates as well 
as the mutual information rate for a time window $T = 32$ ms and time 
resolution $\Delta\tau = 2$ ms. The error bars on the raw entropy and 
information rates were estimated to be approximately $\pm 0.2$ bits/sec 
using a simple bootstrap procedure over the repeated trials.
The extrapolation to infinite data size 
is shown for the mutual information rate estimate (error bars in 
the extrapolated values will be $< \pm 0.2$ bits/sec) and is 
consistent with the 
prediction of Eq.~(\ref{extrap}). Since the total entropy is nearly 
extensive and the noise entropy rate decreases with word duration 
due to subextensive 
corrections as described above, the mutual information rate shown in 
Fig.~\ref{inforate_trials} grows with word duration.
We find that there is an upward change in the mutual information 
rate (computed at $\Delta\tau  = 2$ ms and $T = 32$ ms) 
of $\sim 7$\%, in the large $T$ limit.
For simplicity in the following, we shall look at a fixed word 
duration $T = 32$ ms that is in the well-sampled region for all 
time resolutions $\Delta\tau$ considered.  
 
The mutual information rate measures the rate at which 
the spike train removes uncertainty about the stimulus.
However, the mutual information estimate does not depend on identifying 
either the relevant
features of the stimulus or the relevant features of the response, which
is crucial in analyzing the response to such complex stimuli. In this
sense, our estimates of information transmission and efficiency are 
independent of any model for the code, and provide a benchmark against
which such models could be tested. 

\begin{figure}[t]
\centering
\includegraphics[height=4in]{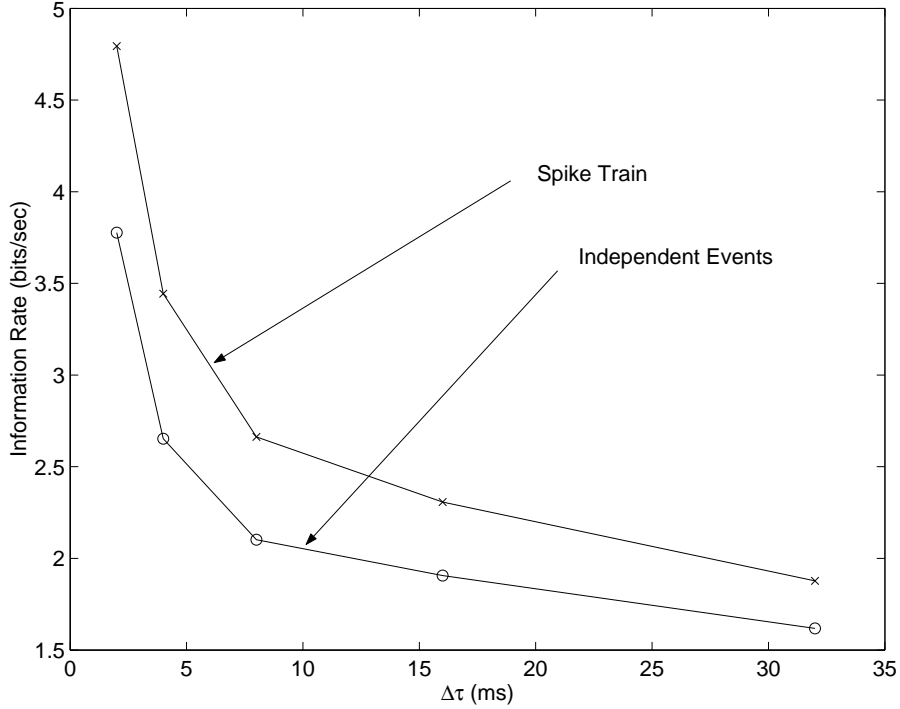}
\caption{Information rates for the spike train ($T = 32$ ms)
and single spike events as a function of time 
resolution $\Delta\tau$ of the spike rasters, corrected for finite 
data size effects.}
\label{inforate_tres}
\end{figure}

One way to look at the information results is to fix our time window $T$
and ask what happens as we change our time resolution $\Delta\tau$.  When
$\Delta\tau = T$, the ``word'' describing the response is nothing but the
number of spikes in the window, so we have a rate or counting code.  As we
decrease $\Delta\tau$, we gradually distinguish more and more detail in
the arrangement of spikes in the window.  We chose a range of $T$ values 
from $30-100$ ms in our analyses to cover previously observed response 
windows for field L neurons and to probe the behaviorally relevant time scale
($\sim 100$ ms) of individual song syllables or notes.  For $T = 32$ ms, we 
show the results (extrapolated to infinite data size) in the upper curve of 
Fig.~\ref{inforate_tres}.
The spike train mutual information shows a clear increase as the timing 
resolution is improved.
In addition, Fig.~\ref{inforate_tres} shows that roughly half of the 
information is accessible
at time resolutions better than $10$ ms and additional information is still
being revealed as time resolution is improved to 2 ms.

\section{Information in rate modulation}

Knowing the mutual information between the stimulus and the spike train 
(defined in the window $T$), we would like to ask whether this can be 
accounted for by the information in single spike events or whether there 
is some additional information conveyed by the patterns of spikes. 
In the latter 
case, we have precisely what we mean by a temporal or timing code:
there is information beyond that attributable to the probability of 
single spike events occurring at time $t$ relative to the onset of the 
stimulus. By event at time $t$, we mean that the event occurs between 
time $t$ and time $t + \Delta\tau$, where $\Delta\tau$ is the resolution 
at which we are looking at the spike train. This probability is simply 
proportional to the firing rate 
(or peri-stimulus time histogram (PSTH)) $r(t)$ at time $t$ normalized by 
the mean firing rate $\bar{r}$. Specifically if the duration of each 
repeated trial is $T_{{\rm repeat}}$ we have 
\begin{equation}
P(1\, \textrm{spk @}\, t | s(t')) = \frac{r(t)\,\Delta\tau}{\bar{r}\, T_{{\rm repeat}}}~,
\end{equation}
where $s(t')$ denotes the stimulus history ($t' < t$).
The probability of a spike event at $t$, {\it a priori} of knowing the 
stimulus history, is flat: $P(1\, \textrm{spk @}\, t) = \Delta\tau/T_{{\rm repeat}}$. Thus, the mutual information between the stimulus and the single spike 
events is~\cite{brenner2000}:
\begin{eqnarray}
I(\mathrm{1\ spike}; \Delta\tau) &=& S[P(1\, \textrm{spk @}\, t)] -  
\left< S[P(1\, \textrm{spk @}\, t | s)]\right>_s \nonumber\\
%&{}& \nonumber\\
&=&\bigg\langle \frac{r(t)}{\bar{r}} 
\log_2\left(\frac{r(t)}{\bar{r}}\right) \bigg\rangle_t \,{\rm bits,}
\label{infoonesp}
\end{eqnarray}
where $r(t)$ is the PSTH binned to resolution $\Delta\tau$ and the 
stimulus average in the first expression is replaced by a time average 
in the second (as
discussed in the calculation of the noise entropy in spike train words 
in the previous section). 
We find that this information is approximately 
$1$ bit for $\Delta\tau = 2$ ms. Supposing that the individual spike
events are independent ({\it i.e.} no intrinsic spike train correlations),
the information {\it rate} in single spike events is obtained by 
multiplying the mutual information per spike (Eq.~\ref{infoonesp}) by 
the mean firing rate of the neuron ($\sim 3.5$ Hz).  This gives an upper 
bound to the single spike event contribution to the information rate and is 
shown in the lower curve of Fig.~\ref{inforate_tres} 
(error bars are again $< \pm 0.2$ bits/sec). 
Comparing with the spike train information (upper curve),
we see that at a resolution 
of $\Delta\tau = 2$ ms, there is at least $\sim 20$\% of the total 
information in the spike train that cannot be attributable to single 
spike events. Thus there is some pattern of spikes that is contributing 
synergistically to the mutual information.
The fact discussed, in the previous section, that the 
spike train information rate {\it grows} 
subextensively with the the word duration out to the point where data sampling becomes problematic is further confirmation of 
the synergy from spike patterns. 
Thus we have shown model-independent 
evidence for a temporal code in the neural responses.

\section{Conclusion}

Until now, few experiments on neural responses in high level,
central vertebrate brain areas have measured the information that these
responses provide about dynamic, naturalistic sensory signals. As
emphasized in earlier work on invertebrate systems, information
theoretic approaches have the advantage that they require no assumptions
about the features
of the stimulus to which neurons respond. Using this method in
the songbird auditory forebrain, we found that patterns of spikes seem to
be special events in the neural code of these neurons, since they carry
more information than expected by adding up the contributions of
individual spikes. It remains to be determined what these spike patterns
are, what stimulus features they may encode, and what mechanisms may be 
responsible for reading such codes at
even higher levels of processing.

\section*{Acknowledgments}

Work at UCSF was supported by grants from the NIH (NS34835) and the 
Sloan-Swartz Center for
Theoretical Neurobiology. BDW and KS supported by NRSA grants from the NIDCD.
We thank Katrin Schenk and Robert Liu for useful discussions.

%\subsubsection*{References}

\end{document}